\documentclass[twoside,11pt]{article}
\usepackage{subfigure}
\usepackage{graphicx}
\usepackage{amsmath}
\usepackage{amssymb}
\begin{document}
\setcounter{page}{1}
\thispagestyle{empty}

\vspace*{-.8in}
\noindent
{\footnotesize {\it Journal of the Calcutta Mathematical Society} (in press)} \\

\begin{center}
{\bf PLANCK'S BLACKBODY RADIATION LAW : PRESENTATION IN DIFFERENT DOMAINS AND DETERMINATION OF THE RELATED DIMENSIONAL CONSTANTS} \\
\vspace{1em}
{\bf GERHARD KRAMM$^1$, and NICOLE M\"{O}LDERS$^{1,2}$} \\
\end{center}


\noindent {\footnotesize {\bf Abstract.} In this paper the Planck
function is derived in the frequency domain using the method of
oscillators. It is also presented in the wavelength domain and in
the wave number domain. The latter is mainly used in spectroscopy
for studying absorption and emission by gases. Also the power law of
Stefan and  Boltzmann is derived for these various domains. It is
shown that this power law is generally independent of the domain in
which the Planck function is presented. A formula for the filtered
spectrum is also given and expressed in the sense of the power law
of Stefan and Boltzmann. Furthermore, based on Wien's displacement
relationship, it is argued that the wavelength at which the maximum
of the Planck function presented in the wavelength domain occurs
differs from that of the maxima of the other domains by a factor of
$1.76$. As Planck determined his elementary quantum of action,
eventually called the Planck constant, and the Boltzmann constant
using Wien's displacement relationship formulated for the wavelength
domain, it is shown that the values of these fundamental constants
are not affected by the choice of domain in which the Planck
function is presented. Finally, the origin of the Planck constant is
discussed.}

\section{Introduction}

In 1999 Soffer and Lynch [1999] discussed the color sensitivity of the eye in terms of the spectral properties and the photo and chemical stability of available biological materials. Based on Planck's blackbody radiation law formulated for the wavelength domain and the frequency domain they already illustrated that the peak brightness of the solar spectrum is located in the green range when plotted in wavelength units, but in the near-infrared range when plotted in frequency units. Soffer and Lynch [1999], therefore, stated that the often-quoted notion that evolution led to an optimized eye whose sensitivity peaks where there is most available sunlight is misleading and erroneous. They found that the eye does not appear to be optimized for detection of the available sunlight, including the surprisingly large amount of infrared radiation in the environment.

The intensity of blackbody radiation may not only be plotted in wavelength and frequency units, but also in units of wave numbers because the wave number domain is used in the field of spectroscopy for studying the absorption and emission by gases (Kidder and Vonder Haar [1995]). Students of courses on atmospheric radiation, for instance, often have difficulties to understand why the maximum of Planck's radiation law presented in the wave number domain differs from that located in the wavelength domain (see Figure 1). In doing so it is important to realize that wave numbers used in spectroscopy differ from those considered in physics by a factor of $2\;\pi$.

Since Planck [1901] considered - beside the velocity of light in vacuum - Stefan's constant for estimating the ratio $k^4/h^3$ and Wien's [1894] displacement relationship $\lambda_{max}^{(\lambda)}\;T = const.$, for determining the ratio $h/k = const.$, it is indispensable to show that his way to obtain values for his "Wirkungsquantum", $h$, (this means an elementary quantum of action eventually called the Planck constant), and the Boltzmann constant, $k$, is independent of the domain in which Planck's radiation law is presented. Here, $\lambda_{max}^{(\lambda)}$  is the wavelength at which the maximum of Planck's radiation formula occurs in the wavelength domain, and $T$ is the absolute temperature.

Planck [1901] introduced an element of energy (see his equation (4)) eventually related to $h\;\nu$ (see his §10), where $\nu$ is the frequency. This means that Planck assumed that energy is quantized, but not light, as often misquoted. A derivation of Planck's radiation law clearly demonstrates this fact. The quantization of light was eventually introduced by Einstein [1905] when he related the monochromatic radiation to mutually independent light quanta (or photons) and their magnitude, in principle, to $h\; \nu$. Einstein even did not use Planck's radiation law in its exact manner, but its approximation that fits the exponential law of Wien [1896] and Paschen [1896].

Planck's radiation function also plays an important role as the only
source function in the radiative transfer equation when a
non-scattering medium is in local thermodynamic equilibrium so that
a beam of monochromatic intensity passing trough the medium will
undergo absorption and emission processes simultaneously, as
described by Schwarzschild's equation (Liou [2002], Chandrasekhar
[1960], Goody \verb"&" Yung [1989], Lenoble [1993]). This is, for
instance, the case of the transfer of infrared radiation emitted by
the earth and its atmosphere. This kind of radiation plays a
prominent role in the physics of climate (Liou [2002], Peixoto
\verb"&" Oort [1992]).Therefore, a derivation of this radiation law
is very helpful to understand its physical background.

There are various ways to derive Planck's radiation law. Beside
Planck's original one (that did not fully satisfy him), different
derivations were published, for instance, by Einstein [1917], Bose
[1924], Pauli [1973], and Kramm and Herbert [2006]. Einstein's
derivation of Planck's blackbody radiation law is described, for
instance, in the textbooks of Feynman {\it et al.} [1963], Semat and
Albright [1972], Liou [2002], and Rybicki and Lightman [2004]. In a
very instructive and simplified derivation Einstein introduced the
important concept regarding the probability of the emission
(including stimulated emission) and/or absorption of radiation.
Bose's [1924] derivation is based on a consequent application of the
Boltzmann [1877] statistics. Kramm and Herbert [2006] used
principles of dimensional analysis in their heuristic derivation of
Planck's law. They showed that a dimensional constant eventually
identified as Planck's constant is required to establish a
similarity function indispensable to avoid the Rayleigh-Jeans
catastrophe in the ultraviolet (Ehrenfest [1911]).

\begin{figure}
\begin{center}
\includegraphics[width=0.6\textwidth]{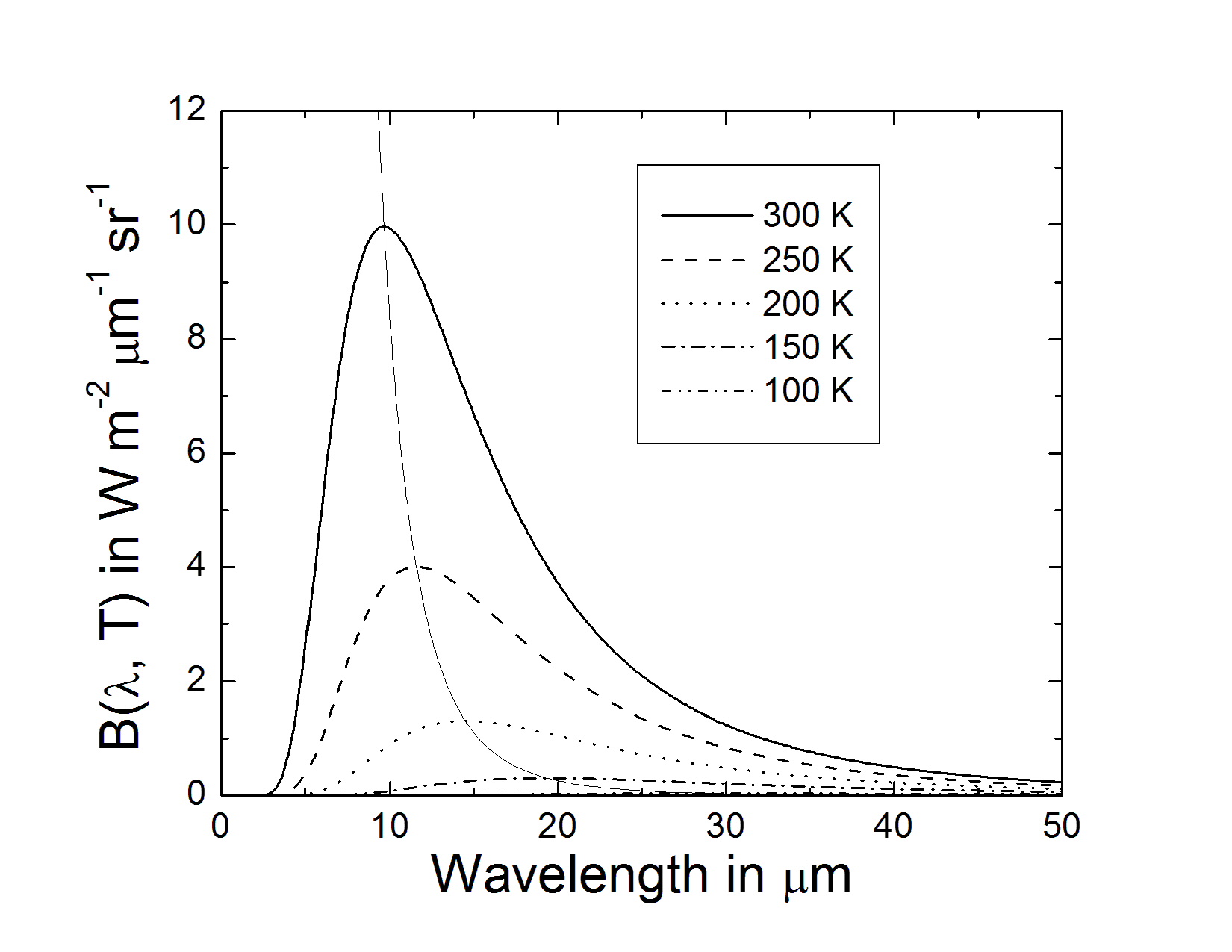}

\includegraphics[width=0.6\textwidth]{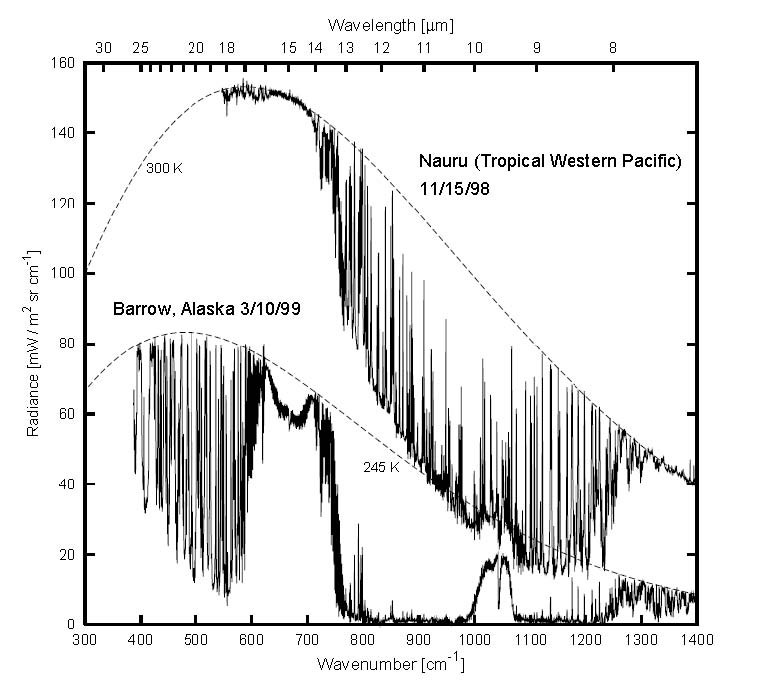}
\end{center}
\vspace*{1em}
\noindent
Fig. 1. The Planck function illustrated for different temperatures in the wavelength
domain (above) and in the wave number domain (below). The latter, adopted from
Petty [2004], shows two examples of measured emission spectra together with the
Planck functions corresponding to the approximate surface temperatures (dashed
lines). (Data courtesy of Robert Knuteson, Space Science and Engineering Center,
University of Wisconsin-Madison.)
\end{figure}

\newpage
The following derivation of the Planck function in the frequency
domain, presented in Section 2, is linked to various sources (Planck
[1901], Pauli [1973], D\"{o}ring [1973], Landau \verb"&" Lifshitz
[1980], Rybicki \verb"&" Lightman [2004]). This derivation is called
the method of oscillators (Pauli [1973]). In Sections 3 and 4 the
Planck function is presented in the wavelength domain and in the two
wave number domains. The derivation of the power law of Stefan
[1879] and Boltzmann [1884] is presented in Section 5. It is shown
that this power law is generally independent of the domain in which
the Planck function is presented. Also a formula for the filtered
spectrum is derived and expressed in the sense of the power law of
Stefan and Boltzmann. In Section 6, Wien's displacement relationship
is derived for the various domains. It is shown that
$\lambda_{max}^{(\lambda)}$ differs from $\lambda_{max}^{(\nu)}$ of
the frequency domain and $\lambda_{max}^{(n_s)}$ and
$\lambda_{max}^{(n_p)}$ of the two wave number domains used in
spectroscopy (superscript $n_s$) and physics (superscript $n_p$) by
a factor of $1.76$. Fortunately, the values of the fundamental
constants, $h$ and $k$, are not affected by the choice of domain in
which the Planck function is presented. The origin of the Planck
constant is discussed in Section 7.
\section{The Planck function in the frequency domain}

An example of a perfect blackbody radiation is the "Hohlraumstrahlung" that describes the radiation in a cavity bounded by any emitting and absorbing opaque substances of uniform temperature. According to Kirchhoff's [1860] findings, the state of the thermal radiation in such a cavity is entirely independent of the nature and properties of these substances and only depends on the absolute temperature, $T$, and the frequency, $\nu$ (or the radian frequency, $\omega = 2\;\pi\;\nu$ or the wavelength, $\lambda$). The radiation that ranges from $\nu$ to $\nu + d\nu$ contributes to the field of energy within a volume $dV$, on average, an amount of energy that is proportional to $d\nu$ and $dV$ expressed by (e.g., D\"{o}ring [1973], Landau \verb"&" Lifshitz [1980])

\begin{figure}
\begin{center}
\includegraphics[width=0.6\textwidth]{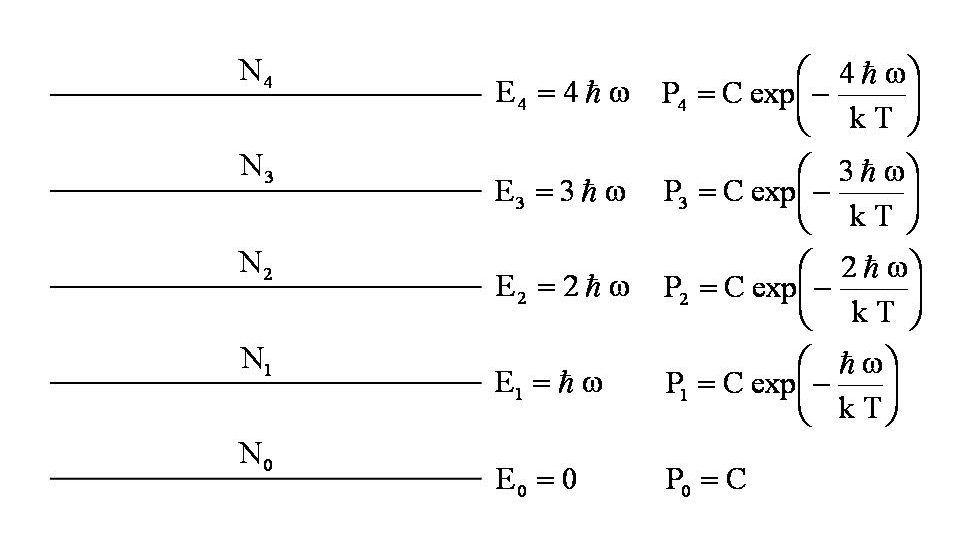}

\end{center}
\vspace*{1em}
\noindent
Fig. 2. The energy levels of a harmonic oscillator are equally spaced by $\Delta E = {E_{n + 1}}\; - \;{E_n} = \hbar\; \omega$ for $n = 0,\;1,\;2,\; \ldots ,\;\infty $ (adopted from Feynman {\it et al.} [1963]).
\end{figure}

\begin{equation}
dE = U\left( {\nu ,\;T} \right)\;d\nu \;dV = U\left( {\omega ,\;T}
\right)\;d\omega \;dV\quad.
\end{equation}
The quantity $U\left( {\nu ,\;T} \right)$ (or $U\left( {\omega ,\;T} \right)$) is called the monochromatic (or spectral) energy density of radiation. According to Planck, in the case of thermal equilibrium, it may be related to the average energy, $\overline E $, of a harmonic oscillator of the frequency $\nu$ located inside the cavity walls by (Planck [1901], Pauli [1973])
\begin{equation}
U\left( {\nu ,\;T} \right) = A\;\overline E\quad,
\end{equation}
where $A$ is a constant. The quantities $A$ and $\overline E $ have to be determined.

In the case of the thermal equilibrium, the probability, $P\left( {{E_j}} \right)$, to detect a stationary state with an energy ${E_j}$ is given by (Boltzmann [1877])
\begin{equation}
P\left( {{E_j}} \right) = \alpha \;{g_j}\;\exp \left( { -
\;\frac{{{E_j}}} {{k\;T}}} \right)\quad.
\end{equation}

\noindent Here, $\alpha$ is a constant, $g_j$ is the number of stationary states, and $k = 1.3806 \cdot {10^{\, - \;23}}\;J\;{K^{\, - \;1}}$ is the Boltzmann constant. Equation (3) reflects Boltzmann's connection between entropy and probability. Analogous to Boltzmann's formula, we express the probability that a harmonic oscillator occupies the $n^{th}$ level of energy, ${E_n}$, by
\begin{equation}
{P_n} = P\left( {{E_n}} \right)\; = \;C\;\exp \left( { -
\;\frac{{{E_n}}} {{k\;T}}} \right)\quad,
\end{equation}
\noindent where $C$ is another constant. Planck [1901] postulated that this amount of energy is given by
\begin{equation}
{E_n} = n\;h\;\nu = n\;\hbar \;\omega
\end{equation}
\noindent which, in principle, means that the energy is quantized. (From a historical point of view, his postulate can be considered as the initiation quantum physics.) Here, $n = 0,\;1,\;2,\; \ldots ,\;\infty $, is an integer, the so-called quantum number, $h = 6.626 \cdot {10^{ - 34}}\;J\;s$ is the Planck constant, and $\hbar = h /\left( 2\;\pi\right)$ is the Dirac constant. Planck assumed that the energy of an oscillator in the ground state ($n = 0$) equals zero. Today we know that for $n = 0$ the zero energy is given by $E_0 = 1/2\;h\;\nu$ so that Eq. (5) becomes (e.g., D\"{o}ring [1973], Pauli [1973], Liou [2002])
\begin{equation}
{E_n} = \left( {n\; + \;\frac{1} {2}} \right)\;h\;\nu  = \left( {n\;
+ \;\frac{1} {2}} \right)\;\hbar \;\omega \quad.
\end{equation}
\noindent However, the fact stated in Eq. (6) does not notably affect the validity of Planck's conclusions. Furthermore, he postulated that the quanta of energy are only emitted when an oscillator changes from one to another of its quantized energy states according to $\Delta E = {E_{n + 1}}\; - \;{E_n} = h\;\nu = \hbar \;\omega $ for $n = 0,\;1,\;2,\; \ldots ,\;\infty $. This value is called a quantum of energy. \indent Obviously, the constant $C$ occurring in Eq. (4) can be determined from the condition that the sum over all probabilities must be equal to unity, i.e.,
\begin{equation}
\sum\limits_{n\; = \;0}^\infty  {{P_n}} = \sum\limits_{n\; =
\;0}^\infty  {C\;\exp \left( { - \;\frac{{{E_n}}} {{k\;T}}} \right)}
\; = \;C\;\sum\limits_{n\; = \;0}^\infty  {\exp \left( { -
\;\frac{{{E_n}}} {{k\;T}}} \right)}  = 1 \quad.
\end{equation}
\noindent Thus, we have
\begin{equation}
C = \frac{1} {{\sum\limits_{n\; = \;0}^\infty  {\exp \left( { -
\;\frac{{{E_n}}} {{k\;T}}} \right)} }}\quad.
\end{equation}
\noindent Now, we consider a lot of oscillators each being a vibrator of frequency $\nu$. Some of these oscillators, namely $N_0$, will be in the ground state ($n = 0$), $N_1$ oscillators will be in the next higher one ($n = 1$), and so forth (see Figure 2). Thus, at the $n^{th}$ energy level we have an energy amount of ${\varepsilon _n} = {E_n}\;{N_n}$, where the number of harmonic oscillators that occupies this level is related to the corresponding probability by ${N_n} = N\;{P_n}\left( {{E_n}} \right)$. Thus, we have
\begin{equation}
{\varepsilon _n} = {E_n}\;{N_n} = {E_n}\;N\;C\;\exp \left( { -
\;\frac{{{E_n}}} {{k\;T}}} \right) = \frac{{{E_n}\;N\;\exp \left( {
- \;\frac{{{E_n}}} {{k\;T}}} \right)}} {{\sum\limits_{n\; =
\;0}^\infty  {\exp \left( { - \;\frac{{{E_n}}} {{k\;T}}} \right)}
}}\quad.
\end{equation}
\noindent According to
\begin{equation}
N = \sum\limits_{n\; = \;0}^\infty  {{N_n}} = \sum\limits_{n\; =
\;0}^\infty  {N\;C\;\exp \left( { - \;\frac{{{E_n}}} {{k\;T}}}
\right)}  = N\;\underbrace {\sum\limits_{n\; = \;0}^\infty {C\;\exp
\left( { - \;\frac{{{E_n}}} {{k\;T}}} \right)} }_{ =
\;1\;(see\;Eq.\;(7))} = N \quad,
\end{equation}
\noindent we may state that $N$ is the total number of harmonic
oscillators. The total energy is then given by
\begin{equation}
E = \sum\limits_{n\; = \;0}^\infty  {{\varepsilon _n}}  =
\frac{{N\;\sum\limits_{n\; = \;0}^\infty  {{E_n}\;\exp \left( { -
\;\frac{{{E_n}}} {{k\;T}}} \right)} }} {{\sum\limits_{n\; =
\;0}^\infty  {\exp \left( { - \;\frac{{{E_n}}} {{k\;T}}} \right)}
}}\quad.
\end{equation}
\noindent From this equation we can infer that the average energy
per oscillator in thermal equilibrium as required by Eq. (2)
is given by
\begin{equation}
\overline E = \frac{E} {N} = \frac{{\sum\limits_{n\; = \;0}^\infty
{{E_n}\;\exp \left( { - \;\frac{{{E_n}}} {{k\;T}}} \right)} }}
{{\sum\limits_{n\; = \;0}^\infty  {\exp \left( { - \;\frac{{{E_n}}}
{{k\;T}}} \right)} }}\quad.
\end{equation}
\noindent For the purpose of simplicity we set
\begin{equation}
Z = \sum\limits_{n\; = \;0}^\infty  {\exp \left( { -
\;\frac{{{E_n}}} {{k\;T}}} \right)}\quad.
\end{equation}
\noindent The derivative of $Z$ with respect to $T$ amounts to
\begin{equation}
\frac{{dZ}} {{dT}} = \sum\limits_{n\; = \;0}^\infty  {\exp \left( {
- \;\frac{{{E_n}}} {{k\;T}}} \right)\;\left( {\frac{{{E_n}}}
{{k\;{T^2}}}} \right)} = \frac{1} {{k\;{T^2}}}\;\sum\limits_{n\; =
\;0}^\infty  {{E_n}\;\exp \left( { - \;\frac{{{E_n}}} {{k\;T}}}
\right)}
\end{equation}
\noindent or
\begin{equation}
k\;{T^2}\;\frac{{dZ}} {{dT}} = \sum\limits_{n\; = \;0}^\infty
{{E_n}\;\exp \left( { - \;\frac{{{E_n}}} {{k\;T}}} \right)}\quad.
\end{equation}
\noindent Combining Eqs. (12) and (15) yields
\begin{equation}
\overline E  = \frac{{k\;{T^2}}} {Z}\;\frac{{dZ}} {{dT}} =
k\;{T^2}\;\frac{d} {{dT}}\;\left( {\ln \;Z} \right)\quad.
\end{equation}
\noindent As $E_n$ is quantized (see Eq. (5)), we obtain
\begin{equation}
Z = \sum\limits_{n\; = \;0}^\infty  {\exp \left( { -
\;\frac{{n\;h\;\nu }} {{k\;T}}} \right)}  = \sum\limits_{n\; =
\;0}^\infty  {{{\left( {\exp \left( { - \;\frac{{h\;\nu }} {{k\;T}}}
\right)} \right)}^n}}\quad.
\end{equation}
\noindent If we define
\begin{equation}
x = \exp \left( { - \;\frac{{h\;\nu }} {{k\;T}}} \right)\quad,
\end{equation}
\noindent we will easily recognize that
\begin{equation}
Z = \sum\limits_{n\; = \;0}^\infty  {{x^n}}
\end{equation}
\noindent is a geometric series. As $0 \le x < 1$,
its sum is given by
\begin{equation}
Z = \sum\limits_{n\; = \;0}^\infty  {{x^n}} = \frac{1} {{1\; - \;x}}
= \frac{1} {{1\; - \;\exp \left( { - \;\frac{{h\;\nu }} {{k\;T}}}
\right)}}\quad.
\end{equation}
\noindent Inserting this expression into Eq. (16) yields
\begin{eqnarray}
\nonumber \overline E  = k\;{T^2}\;\frac{d} {{dT}}\;\left( {\ln \;Z}
\right) &=& - \;k\;{T^2}\;\frac{d} {{dT}}\;\left\{ {\ln \left( {1\;
- \;\exp \left( { - \;\frac{{h\;\nu }} {{k\;T}}} \right)} \right)}
\right\}\\
\nonumber &=& - \;k\;{T^2}\;\frac{1} {{1\; - \;\exp \left( { -
\;\frac{{h\;\nu }} {{k\;T}}} \right)}}\;\left( { - \;\exp \left( { -
\;\frac{{h\;\nu }} {{k\;T}}} \right)\;\left( {\frac{{h\;\nu }}
{{k\;{T^2}}}} \right)} \right)\\
&=& \frac{{h\;\nu \;\exp \left( { - \;\frac{{h\;\nu }} {{k\;T}}}
\right)}} {{1\; - \;\exp \left( { - \;\frac{{h\;\nu }} {{k\;T}}}
\right)}}
\end{eqnarray}
\noindent or
\begin{equation}
\overline E = \frac{{h\;\nu }} {{\exp \left( {\frac{{h\;\nu }}
{{k\;T}}} \right)\; - \;1}}\quad.
\end{equation}
\noindent Inserting this equation into Eq. (2) provides
\begin{equation}\label{2.23}
U\left( {\nu ,\;T} \right) = A\;\frac{{h\;\nu }} {{\exp \left(
{\frac{{h\;\nu }} {{k\;T}}} \right)\; - \;1}}\quad.
\end{equation}
\noindent The expression
\begin{equation}
\overline n  = \frac{1} {{\exp \left( {\frac{{h\;\nu }} {{k\;T}}}
\right)\; - \;1}} = \frac{1} {{\exp \left( {\frac{{\hbar \;\omega }}
{{k\;T}}} \right)\; - \;1}}
\end{equation}
\noindent is customarily called the Planck distribution. It may be
regarded as a special case of the Bose-Einstein distribution when
the chemical potential of a "gas" of photons is given by $\mu\;=\;0$
(Bose [1924], Einstein [1924], Landau \verb"&" Lifshitz [1980],
Rybicki \verb"&" Lightman [2004]).

Now, we have to determine the constant $A$. It can be
inferred from the classical blackbody radiation law,
\begin{equation}
U\left( {\nu ,\;T} \right) = \frac{{8\;\pi \;{\nu ^2}}}
{{{c^3}}}\;k\;T \quad,
\end{equation}
\noindent where $c = 2.998 \cdot {10^{\,8}}\;m\;{s^{ - \;1}}$ is the
velocity of light in vacuum. This radiation law was first derived by
Rayleigh [1900, 1905] using principles of classical statistics, with
a correction by Jeans [1905]. Today it is called the Rayleigh-Jeans
law. Note that Lorentz [1903] derived it in a somewhat different
way. This classical radiation law fulfills both (a) Kirchhoff's
findings regarding the state of the thermal radiation in a cavity,
and (b) the requirements of Wien's [1894] displacement law that
reads (e.g., Planck [1901], Ehrenfest [1911], Pauli [1973])
\begin{equation}
U\left( {\nu,\;T} \right) \propto {\nu ^3} f\left(\frac{\nu}{T}
\right)\quad.
\end{equation}
\noindent For small frequencies at relatively high temperature,
formula (25) works well. It was experimentally proved by
Lummer and Pringsheim [1900] and Rubens and Kurlbaum [1900, 1901].
Thus, Planck [1901] already stated that the law of the energy
distribution within the normal spectrum derived by Wien [1896] on
the basis of molecular kinetic considerations (experimentally proved
by Paschen [1896]) and later deduced by himself on the basis of the
theory of the electromagnetic radiation and the $2^{nd}$ law of
thermodynamics (Planck [1900a, 1900b]) cannot be generally valid.
The Rayleigh-Jeans law, of course, cannot be correct for high values
of $\nu$ because for $\nu \; \to \;\infty$ the monochromatic energy
density, $U\left(\nu,\;T \right)$, would tend to infinity (Einstein
[1905]). Ehrenfest [1911] coined this behavior the Rayleigh-Jeans
catastrophe in the ultraviolet. Therefore, we consider Planck's
formula in the red range for which the Rayleigh-Jeans law is valid.
This consideration is related to Ehrenfest's [1911] red requirement.

For $\nu \to 0$  Eq. (23) provides $U\left(\nu,\;T \right) \to 0/0$.
Thus, we have to use de l'Hospital's rule to determine
$U\left(\nu,\;T \right)$ for this limit. Defining $f\left(\nu\right)
= A\;h\;\nu$ and $g\left(\nu\right) = \exp\left(
h\;\nu/\left(k\;T\right)\right)\; - \;1$ leads to
\begin{equation}
\lim_{\nu \; \to \;0} \;\frac{{f'\left( \nu  \right)}} {{g'\left(
\nu  \right)}} = \lim_{\nu \; \to \;0} \;\frac{{A\;h}} {{\frac{h}
{{k\;T}}\;\exp \left( {\frac{{h\;\nu }} {{k\;T}}} \right)\;}} =
A\;k\;T \quad.
\end{equation}
\noindent Comparing Eqs.(25) and (27) yields
\begin{equation}
A = \frac{{8\;\pi \;{\nu ^2}}} {{{c^3}}}\quad,
\end{equation}
\noindent as already mentioned by Planck [1901]. Inserting this
expression into Eq. (23) provides (Planck [1901], Pauli [1973])
\begin{equation}
U\left( {\nu ,\;T} \right) = \frac{{8\;\pi \;h}}
{{{c^3}}}\;\frac{{{\nu ^3}}} {{\exp \left( {\frac{{h\;\nu }}
{{k\;T}}} \right)\; - \;1}}\quad.
\end{equation}
\noindent Thus, Eq. (1) may be written as
\begin{equation}
dE = \frac{{8\;\pi \;h}} {{{c^3}}}\;\frac{{{\nu ^3}}} {{\exp \left(
{\frac{{h\;\nu }} {{k\;T}}} \right)\; - \;1}}\;d\nu \;dV
\end{equation}
\noindent or
\begin{equation}
dE = \frac{\hbar } {{{\pi ^2}\;{c^3}}}\;\frac{{{\omega ^3}}} {{\exp
\left( {\frac{{\hbar \;\omega }} {{k\;T}}} \right)\; -
\;1}}\;d\omega \;dV \quad.
\end{equation}
\indent The monochromatic intensity, $B\left( {\nu ,\;T} \right)$,
is generally related to the differential amount of radiant energy,
$dE$, that crosses an area element, $dA$, in directions confined to
a differential solid angle, $d\Omega$, being oriented at an angle
$\theta$ to the normal of $dA$,
\begin{equation}
dE = B\left( {\nu ,\;T} \right)\;\cos \theta \;dA\;d\Omega \;d\nu
\;dt\quad,
\end{equation}
\noindent in the time interval between $t$ and $t\;+\;dt$ and the
frequency interval between $\nu$ and $\nu\;+\;d\nu$. Thus, we
obtain
\begin{eqnarray}\label{2.33}
\nonumber dE &=& \;\frac{{8\;\pi \;h}} {{{c^3}}}\;\frac{{{\nu ^3}}}
{{\exp \left( {\frac{{h\;\nu }} {{k\;T}}} \right)\; - \;1}}\;d\nu
\;dV = B\left( {\nu ,\;T} \right)\;\cos \theta \;dA\;d\Omega
\;dt\;d\nu \\
&=& \frac{{4\;\pi }} {c}\;B\left( {\nu ,\;T} \right)\;\underbrace
{\cos \theta \;dA\;\frac{{d\Omega }} {{4\;\pi }}\;c\;dt}_{ =
\;dV}\;d\nu  = \frac{{4\;\pi }} {c}\;B\left( {\nu ,\;T}
\right)\;d\nu \;dV
\end{eqnarray}
\noindent and, hence (Chandrasekhar [1960], Liou [2002])
\begin{equation}\label{2.34}
B\left( {\nu ,\;T} \right) = \frac{{2\;h}} {{{c^2}}}\;\frac{{{\nu
^3}}} {{\exp \left( {\frac{{h\;\nu }} {{k\;T}}} \right)\; -
\;1}}\quad.
\end{equation}
\noindent The quantity $d\Omega/(4\;\pi)$ in Eq. (33)
expresses the probability of radiation propagation in a certain
direction. Using the relationship
\begin{equation}
B\left( {\vartheta ,\;T} \right)\;d\vartheta = B\left( {\nu \left(
\vartheta  \right),\;T} \right)\;d\nu \quad,
\end{equation}
\noindent where $\vartheta$ stands for any variable like radian
frequency, $\omega$, wavelength, $\lambda$, wave number as defined
in spectroscopy, $n_s = 1/\lambda = \nu/c$, or the wave number as
defined in physics $n_p = 2\;\pi /\lambda = 2\;\pi \;\nu/c = \omega
/c$, that can be related to the frequency $\nu$ via the
transformation $\nu \left(\vartheta \right)$, yields then
\begin{equation}
B\left( {\omega ,\;T} \right) = B\left( {\nu \left( \omega
\right),\;T} \right)\;\frac{{d\nu }} {{d\omega }} = \frac{\hbar }
{{4\;{\pi ^3}\;{c^2}}}\;\frac{{{\omega ^3}}} {{\exp \left(
{\frac{{\hbar \;\omega }} {{k\;T}}} \right)\; - \;1}}\quad.
\end{equation}
\noindent Equations (34) and (36) are customarily
called the Planck functions for these two frequency domains.
\section{The Planck function in the wavelength domain}

\noindent The frequency domain is given by $[0,\;\infty]$. As
\begin{equation}
B\left( {\lambda ,\;T} \right) = B\left( {\nu \left( \lambda
\right),\;T} \right)\;\frac{{d\nu }} {{d\lambda }} = - \;B\left(
{\nu \left( \lambda  \right),\;T} \right)\;\frac{c} {{{\lambda
^2}}}\quad,
\end{equation}
\noindent we obtain for the Planck function in the wavelength domain
$[\infty,\;0]$
\begin{equation}
B\left( {\lambda ,\;T} \right) = - \;\frac{{2\;h\;{c^{\,2}}}}
{{{\lambda ^5}\;\left\{ {\exp \left( {\frac{{h\;c}} {{\lambda
\;k\;T}}} \right)\; - \;1} \right\}}}\quad.
\end{equation}
\section{The Planck function in the wave number domain}
\noindent Since the wave numbers defined in spectroscopy by $n_s =
1/\lambda = \nu/c$ and in physics by $n_p = 2\;\pi /\lambda = 2\;\pi
\;\nu/c = \omega/c$ differ from each other by the factor of
$2\;\pi$, we obtain two slightly different results:

Spectroscopy:
\begin{equation}
B\left( {{n_s},\;T} \right) = B\left( {\nu \left( {{n_s}}
\right),\;T} \right)\;\frac{{d\nu }} {{d{n_s}}} =
2\;h\;{c^{\,2}}\;\frac{{{n_s}^3}} {{\exp \left(
{\frac{{h\;c\;{n_s}}} {{k\;T}}} \right)\; - \;1}}\quad.
\end{equation}
\indent Physics:
\begin{equation}
B\left( {{n_p},\;T} \right) = B\left( {\nu \left( {{n_p}}
\right),\;T} \right)\;\frac{{d\nu }} {{d{n_p}}} = \frac{{2\;\hbar
\;{c^{\,2}}}} {{{{\left( {2\;\pi }
\right)}^{\,3}}}}\;\frac{{{n_p}^3}} {{\exp \left( {\frac{{\hbar
\;c\;{n_p}}} {{k\;T}}} \right)\; - \;1}}\quad.
\end{equation}
\noindent Equation (39) is illustrated in the lower part of
Figure 1 for two different surface temperatures
together with two examples of measured emission spectra.

\section{The power law of Stefan and Boltzmann}
\noindent Integrating Eq. (34) over all frequencies yields
for the total intensity
\begin{equation}
B\left( T \right) = \int\limits_0^\infty  {B\left( {\nu ,\;T}
\right)} \;d\nu = \frac{{2\;h}} {{{c^2}}}\;\int\limits_0^\infty
{\frac{{{\nu ^3}}} {{\exp \left( {\frac{{h\;\nu }} {{k\;T}}}
\right)\; - \;1}}\;d\nu}\quad.
\end{equation}
\noindent Defining $X = h\;\nu / \left(k\;T\right)$ leads to
\begin{equation}
B\left( T \right) = \int\limits_0^\infty  {B\left( {\nu ,\;T}
\right)} \;d\nu  = \frac{{2\;{k^4}}}
{{{c^2}\;{h^3}}}\;{T^4}\;\int\limits_0^\infty  {\frac{{{{X}^3}}}
{{\exp \left( {X} \right)\; - \;1}}\;d{X}}\quad,
\end{equation}
\noindent where the integral amounts to (e.g., Planck [1901], Liou
[2002], Kramm \verb"&" Herbert [2006])
\begin{equation}
\int\limits_0^\infty  {\frac{{{{X}^3}}} {{\exp \left( {X} \right)\;
- \;1}}\;} \;d{X} = \frac{{{\pi ^4}}} {{15}}\quad.
\end{equation}
\noindent Thus, Eqs. (42) and (43) provide for the
total intensity
\begin{equation}
B\left( T \right) = \beta \;{T^4}
\end{equation}
\noindent with
\begin{equation}
\beta = \frac{{2\;{\pi ^4}\;{k^4}}} {{15\;{c^2}\;{h^3}}}\quad.
\end{equation}
\noindent Since blackbody radiance may be considered as an example
of isotropic radiance, we obtain for the radiative flux density also
called the irradiance (see Appendix A)
\begin{equation}
F\left( T \right) = \pi \;B\left( T \right) = \pi \;\beta
\;{T^{\,4}}
\end{equation}
\noindent or
\begin{equation}
F\left( T \right) = \sigma \;{T^4}\quad,
\end{equation}
\noindent where $\sigma = \pi \;\beta = 5.67 \cdot {10^{\, -
\;8}}\;J\;{m^{\, - \;2}}\;{s^{\, - \;1}}\;{K^{\, - \;4}}$ is the
Stefan constant. According to Stefan's [1879] empirical findings and
Boltzmann's [1884] thermodynamic derivation, formula (47) is
customarily called the power law of Stefan and Boltzmann.

\indent On the other hand, the integration of Eq (38) over
all wavelengths yields (e.g., Chandrasekhar [1960], Liou [2002])
\begin{eqnarray}
\nonumber B\left( T \right) &=& \int\limits_\infty ^0 {B\left(
{\lambda ,\;T} \right)} \;d\lambda = - \;\int\limits_\infty ^0
{\frac{{2\;h\;{c^2}}} {{{\lambda ^5}\;\left\{ {\exp \left(
{\frac{{h\;c}} {{\lambda \;k\;T}}} \right)\; - \;1}
\right\}}}\;d\lambda }\\
&=& \int\limits_0^\infty {\frac{{2\;h\;{c^2}}} {{{\lambda
^5}\;\left\{ {\exp \left( {\frac{{h\;c}} {{\lambda \;k\;T}}}
\right)\; - \;1} \right\}}}\;d\lambda }\quad.
\end{eqnarray}
\noindent Using the definition $X = h\;c/\left(\lambda
\;k\;T\right)$ leads, again, to
\begin{equation}
B\left( T \right) = \frac{{2\;{k^4}}}
{{{c^2}\;{h^3}}}\;{T^4}\;\underbrace {\int\limits_0^\infty
{\frac{{{{X}^3}}} {{\exp \left( {X} \right)\; - \;1}}\;dx} }_{ =
\frac{{{\pi ^4}}} {{15}}\;(see\;Eq.\;(43))} = \frac{{2\;{\pi
^4}\;{k^4}}} {{15\;{c^2}\;{h^3}}}\;{T^{\,4}} = \beta
\;{T^{\,4}}\quad,
\end{equation}
\noindent i.e., Eq. (44) deduced in the frequency domain and
Eq. (49) deduced in the wavelength domain provide identical
results, and, and hence, $k^4/\left(c^2\;h^3\right) =
15\;\sigma/\left(2\;\pi^5\right) = const.$ The same is true for the
two wave number domains. If we consider the transformation from the
frequency domain to any $\vartheta$ domain, where $\vartheta$ stands
for $\omega$, $\lambda$, $n_s$, or $n_p$, we have to recognize that
$B\left( {\vartheta ,\;T} \right)\;d\vartheta = B\left( {\nu \left(
\vartheta \right),\;T} \right)\;d\nu$ (see Eq. (35)). This
relationship was already used in Eqs. (36), (37),
(39), and (40). On the other hand, according to the
theorem for the substitution of variables in an integral, we have
\begin{equation}
\int\limits_{{\nu _1}}^{{\nu _2}} {B\left( {\nu ,\;T} \right)}
\;d\nu = \int\limits_{{\vartheta _1}}^{{\vartheta _2}} {B\left( {\nu
\left( \vartheta  \right),\;T} \right)} \;\frac{{d\nu \left(
\vartheta  \right)}} {{d\vartheta }}d\vartheta \quad,
\end{equation}
\noindent where $\nu \left( \vartheta  \right)$ is the
transformation from $\nu$ to $\vartheta$, ${\nu _1} = \nu \left(
{{\vartheta _1}} \right)$ , and ${\nu _2} = \nu \left( {{\vartheta
_2}} \right)$. The results of these integrals are identical because
any variable substitution does not affect the solution of an
integral. Thus, the solution is independent of the domain.
Consequently, the ratio $k^4 /h^3$ of the two natural constants can
be derived using the Stefan constant and the velocity of light in
vacuum. One obtains $k^4/h^3 = 1.248 \cdot 10^8\;J\;s^{-\;3}\;K^{-
\;4}$. Note that Planck [1901] found $k^4/h^3 = 1.168 \cdot
10^8\;J\;s^{-\;3}\;K^{- \;4}$, i.e., a value being
only $6.4 \; \verb"%"$ smaller than the current one. If one of these two natural
constants is known, the other one can be determined, too. In accord
with Planck [1901], we consider Wien's [1894] displacement
relationship to finally determine these constants.

\indent As recently pointed out by Gerlich and Tscheuschner [2009],
the Stefan constant is not a universal constant of physics.
Obviously, this $\sigma$ depends on the quantity $\beta$. If we
consider, for instance, Rayleigh's [1900] radiation formula
\begin{equation}
U\left( {\nu ,\;T} \right) = \frac{{8\;\pi \;{\nu ^2}}}
{{{c^3}}}\;k\;T\;\exp \left( { - \;\frac{{h\;\nu }} {{k\;T}}}
\right)
\end{equation}
\noindent which satisfies - like the Planck function - Ehrenfest's [1911] requirements in the red and violet ranges (for details, see Kramm \verb"&" Herbert [2006]), the monochromatic intensity will read
\begin{equation}
B\left( {\nu ,\;T} \right) = \frac{{2\;{\nu ^2}}}
{{{c^2}}}\;k\;T\;\exp \left( { - \;\frac{{h\;\nu }} {{k\;T}}}
\right)\quad.
\end{equation}
\noindent Thus, the total intensity is given by
\begin{equation}
B\left( T \right) = \int\limits_0^\infty  {B\left( {\nu ,\;T}
\right)} \;d\nu  = \frac{{2\;{k^4}}}
{{{c^2}\;{h^3}}}\;{T^4}\;\int\limits_0^\infty  {{{X}^2}\;\exp \left(
{ - \;{X}} \right)\;d{X}} = \frac{{2\;{k^4}}}
{{{c^2}\;{h^3}}}\;{T^4}\;\Gamma \left( 3 \right) = {\beta
_R}\;{T^4}\quad,
\end{equation}
\noindent where $X = h\;\nu/\left(k\;T\right)$ and  ${\beta _R} =
4\;k^4 /\left(c^2\;h^3\right)$. Here, the subscript $R$
characterizes the value deduced from Rayleigh's [1900] radiation
formula. This value differs from that obtained with the Planck
function by a factor, expressed by ${{{\beta _R} = 30\;\beta }
\mathord{\left/{\vphantom {{{\beta _R} = 30\;\beta } {{\pi ^4}}}}
\right. \kern-\nulldelimiterspace} {{\pi ^4}}}\; \cong
\;0.308\;\beta$. Thus, in the case of Rayleigh's radiation formula
the value of the Stefan constant would be given by $\sigma _R =
0.308\;\sigma = 1.75 \cdot 10^{- \;8}\;J\;{m^{- \;2}}\;{s^{-
\;1}}\;{K^{- \;4}}$.

\indent If we consider any finite or filtered spectrum ranging, for
instance, from $\nu_1$ to $\nu_2$, Eq. (42) will become
\begin{equation}
\left. {B\left( T \right)} \right|_{{{X}_1}}^{{{X}_2}} =
\frac{{2\;{k^4}}}
{{{c^2}\;{h^3}}}\;{T^4}\;\int\limits_{{{X}_1}}^{{{X}_2}}
{\frac{{{{X}^3}}} {{\exp \left( {X} \right)\; - \;1}}\;d{\rm X}} =
{\beta _F}\;{T^4}\quad,
\end{equation}
\noindent where the value of the filtered spectrum, $\beta_F$, is
defined by
\begin{equation}
{\beta _F} = \frac{{2\;{k^4}}}
{{{c^2}\;{h^3}}}\;\int\limits_{{{X}_1}}^{{{X}_2}} {\frac{{{{X}^3}}}
{{\exp \left( {X} \right)\; - \;1}}\;d{X}}\quad.
\end{equation}
\noindent This quantity reflects the real world situations (Gerlich
\verb"&" Tscheuschner [2009]). In such a case the Stefan's constant
would become
\begin{equation}
{\sigma _F} = \sigma \;\frac{{15}} {{{\pi
^4}}}\;\int\limits_{{{X}_1}}^{{{X}_2}} {\frac{{{{X}^3}}} {{\exp
\left( {X} \right)\; - \;1}}\;d{X}} = {e_F}\left(
{{{X}_1},\;{{X}_2}} \right)\;\sigma\quad,
\end{equation}
\noindent where the characteristic value of the filtered spectrum,
${e_F}\left( {{{X}_1},\;{{X}_2}} \right)$, is defined by
(e.g., Modest [2003])
\begin{equation}
{e_F}\left( {{{X}_1},\;{{X}_2}} \right) = \frac{{15}} {{{\pi
^4}}}\;\int\limits_{{{X}_1}}^{{{X}_2}} {\frac{{{{X}^3}}} {{\exp
\left( {X} \right)\; - \;1}}\;d{X}}\quad.
\end{equation}
\noindent This means that in the instance of a filtered spectrum the
power law of Stefan and Boltzmann must read
\begin{equation}
{F_F}\left( T \right)\; = \;{e_F}\left( {{{X}_1},\;{{X}_2}}
\right)\;\sigma \;{T^4}\quad.
\end{equation}
\noindent This formula describes the fractional emission of a
blackbody due to a finite or filtered spectrum. Figure 3 illustrates
the quantity ${e_F}\left( {{{X}_1},\;{{X}_2}} \right)$ for different
filtered spectra and various temperatures. Equation (58) has
the form commonly used in the case of gray bodies
that are characterized by imperfect absorption and emission.

\begin{figure}
\begin{center}
\includegraphics[width=0.6\textwidth]{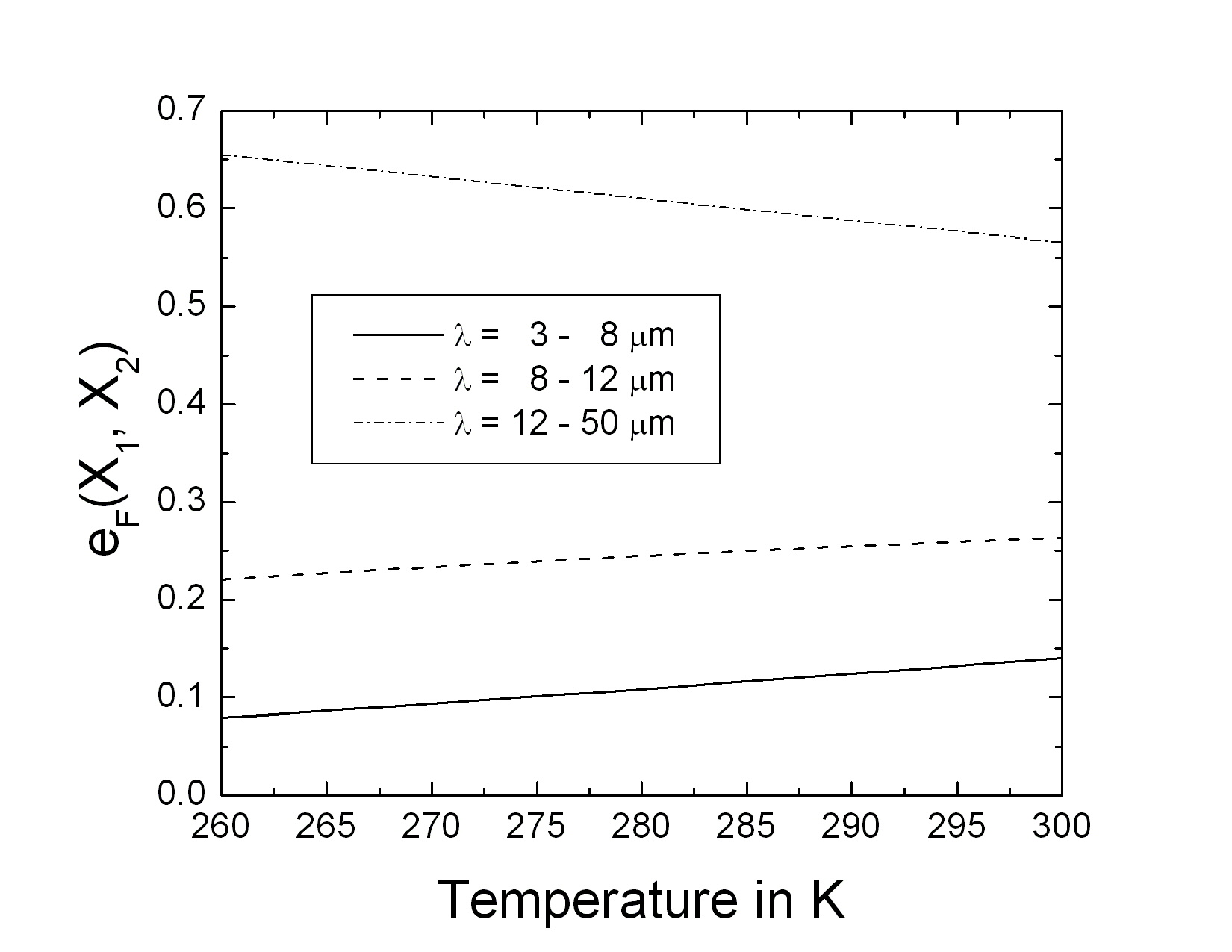}

\end{center}
\vspace*{1em}
\noindent
Fig. 3. Characteristic values, ${e_F}\left( {{{X}_1},\;{{X}_2}} \right)$, of three filtered spectra and various temperatures.
\end{figure}

\section{Wien's displacement relationship}

\noindent Even though Wien's [1894] displacement relationship was
well-known before Planck published his famous radiation law, the
former can simply be derived by determining the maximum of the
latter using both the first derivative test and the second
derivative test. The first derivative of Eq. (34) reads
\begin{equation}
\frac{\partial } {{\partial \nu }}\;B\left( {\nu ,\;T} \right) =
\frac{{2\;h\;{\nu ^{\,2}}}} {{{c^{\,2}}\;{{\left\{ {\exp \left(
{\frac{{h\;\nu }} {{k\;T}}} \right)\; - \;1}
\right\}}^{\,2}}}}\;\left( {3\;\left\{ {\exp \left( {\frac{{h\;\nu
}} {{k\;T}}} \right)\; - \;1} \right\}\; - \;\frac{{h\;\nu }}
{{k\;T}}\;\exp \left( {\frac{{h\;\nu }} {{k\;T}}} \right)}
\right)\quad.
\end{equation}
\noindent This derivative is only equal to zero when the numerator
of the term on the right-hand side of Eq. (59) is equal to
zero (the corresponding denominator is larger than zero for $0 < \nu
< \infty$), i.e.,
\begin{equation}
3\;\left\{ {\exp \left( {{x_\nu }} \right)\; - \;1} \right\}\; =
\;{x_\nu }\;\exp \left( {{x_\nu }} \right)
\end{equation}
\noindent with $x_\nu = h\;\nu/\left(k\;T\right)$. This
transcendental function can only be solved numerically. One obtains
${x_\nu } \cong 2.8214$, and in a further step
\begin{equation}
\frac{{{\nu _{ext}}}} {T} = {x_\nu }\;\frac{k} {h}\quad,
\end{equation}
\noindent where $\nu_{ext}$ is the frequency at which the extreme (either a minimum or a maximum) of the Planck function in the frequency domain occurs. It can simply be proved that for this extreme the second derivative fulfills the condition ${{{\partial ^2}B\left({\nu ,\;T}\right)}/{\partial {\nu ^2} < 0}}$ so that the extreme corresponds to the maximum, $\nu_{max}$. Figure 4 illustrates the value of $\nu_{max}/T$. If we use $c = \lambda \; \nu$ we will obtain

\begin{figure}
\begin{center}
\includegraphics[width=0.6\textwidth]{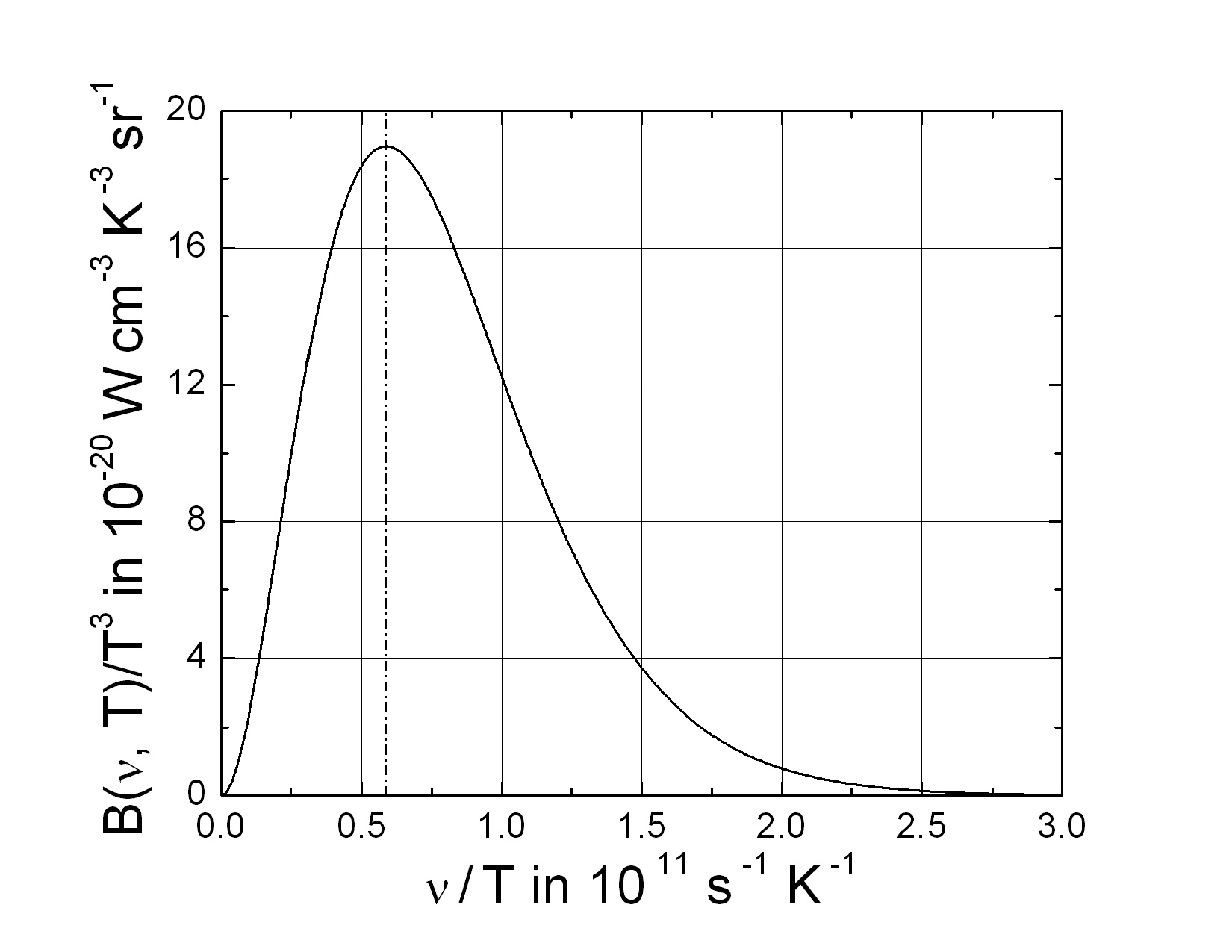}

\end{center}
\vspace*{1em}
\noindent
Fig. 4. The Planck function, as given by Eq. (34), but the curve represents $B\left({\nu ,\;T}\right)/T^3$ versus $\nu/T$ so that it becomes independent of $T$. Note that the constant, $C = 5.879 \cdot {10^{10}}\;{s^{ - \;1}}\;{K^{ - \;1}}$, occurring in Wien's [1894] displacement relationship $C = {\nu_{\max }}/T$ can be inferred from this figure (dashed-dotted line).
\end{figure}

\begin{equation}
\lambda _{\max }^{\left( \nu  \right)}\;T = \frac{c} {{{x_\nu
}}}\;\frac{h} {k} \cong 5.098 \cdot {10^{- \;3}}\;m\;K \quad.
\end{equation}
\noindent This formula should be called Wien's displacement relationship, rather than Wien's displacement law because the latter is customarily used for Eq. (26). For the two different wave number domains, we obtain the same result. Thus, we have
\begin{equation}
\lambda _{\max }^{\left( {{n_p}} \right)} = \lambda _{\max }^{\left(
{{n_s}} \right)} = \lambda _{\max }^{\left( \nu \right)}\quad.
\end{equation}
\noindent On the other hand, if we consider the Planck function (38) formulated for the wavelength domain, the first derivative will read
\begin{equation}
\frac{\partial } {{\partial \lambda }}\;B\left( {\lambda ,\;T}
\right) = \frac{{2\;h\;{c^2}}} {{{\lambda ^{\,6}}{{\left\{ {\exp
\left( {\frac{{h\;c}} {{\lambda \;k\;T}}} \right)\; - \;1}
\right\}}^2}}}\;\left( {5\;\left\{ {\exp \left( {\frac{{h\;c}}
{{\lambda \;k\;T}}} \right)\; - \;1} \right\}\; - \;\frac{{h\;c}}
{{\lambda \;k\;T}}\;\exp \left( {\frac{{h\;c}} {{\lambda \;k\;T}}}
\right)} \right)
\end{equation}
\noindent Again, this derivative is only equal to zero when the numerator of the term on the right-hand side of Eq. (64) equals zero. The corresponding denominator is also larger than zero for $0 < \lambda <\infty$. Thus, defining $x_\lambda = h\;c/\left(\lambda \;k\;T\right)$ leads to (e.g., Planck [1901], Liou [2002], Bohren \verb"&" Clothiaux [2006])
\begin{equation}
5\;\left\{ {\exp \left( {{x_\lambda }} \right)\; - \;1} \right\} =
{x_\lambda }\;\exp \left( {{x_\lambda }} \right)\quad.
\end{equation}
\noindent The numerical solution of this transcendental function reads ${x_\lambda } \cong 4.9651$. Using this result yields
\begin{equation}
\lambda _{\max }^{\left( \lambda  \right)}\;T = \frac{c}
{{{x_\lambda }}}\;\frac{h} {k} \cong 2.897 \cdot {10^{- \;3}}\;m\;K
\quad.
\end{equation}
\noindent The constant at the right-hand-side of this equation can also be taken from Figure 5. Obviously, we have ${{c\;h}/k} = const.$ Note that the requirement ${{{\partial^2}B\left( {\lambda ,\;T} \right)}/{\partial {\lambda^2} < 0}}$ is satisfied for $\lambda_{ext}$ so that this extreme corresponds to a maximum, too.

\begin{figure}
\begin{center}
\includegraphics[width=0.6\textwidth]{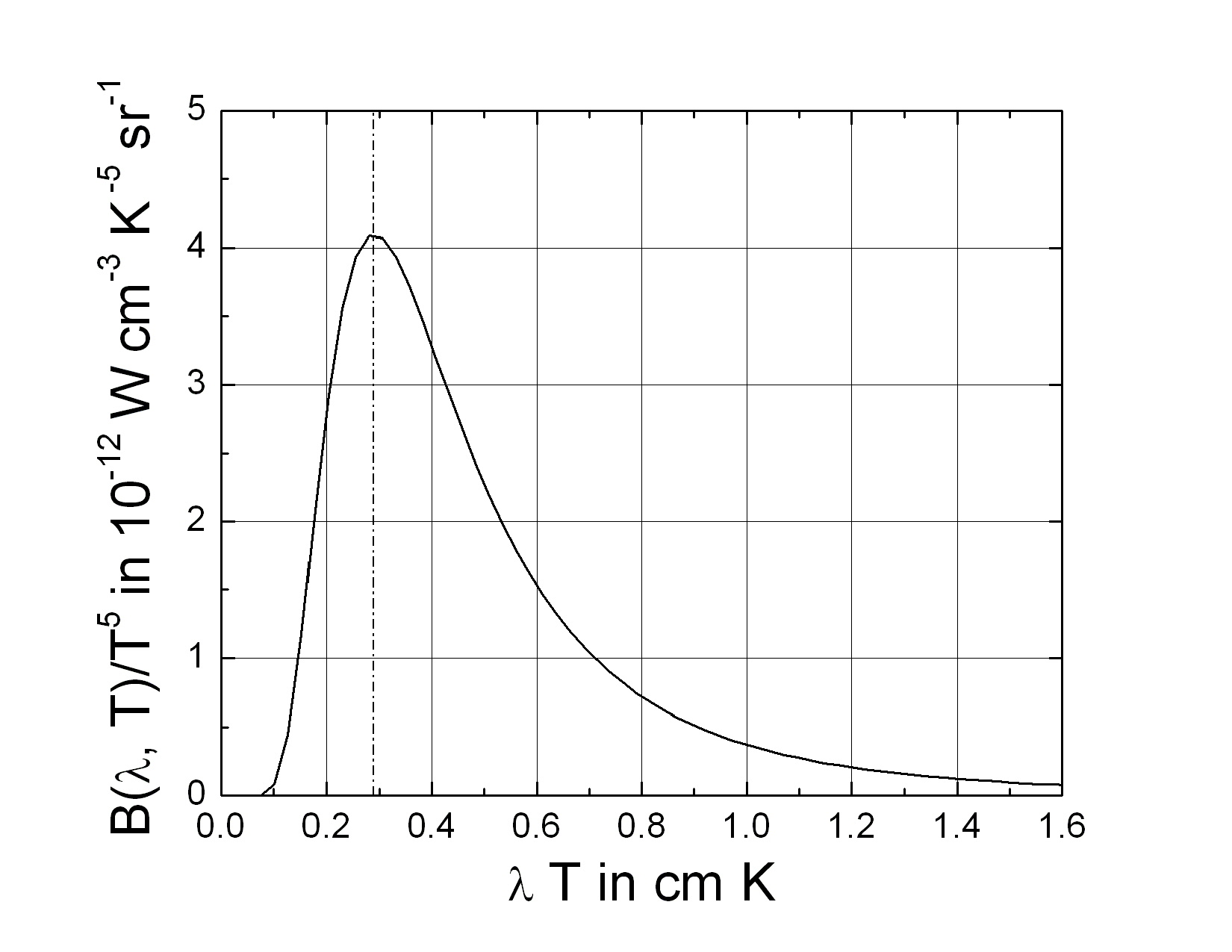}

\end{center}
\vspace*{1em}
\noindent
Fig. 5. The Planck function, as given by Eq. (38), but the curve represents $B\left({\lambda ,\;T}\right)/T^5$ versus $\lambda\;T$ so that it becomes independent of $T$. Note that the constant, $C = 0.2897\;cm\;K$, occurring in Wien's displacement relationship $C = {\lambda_{\max }}\;T$ can be inferred from this figure (dashed-dotted line).
\end{figure}

\indent As expressed by Eqs. (62) and (66), Wien's displacement relationship, derived for the frequency domain and the wave number domains on the one hand and the wavelength domain on the other hand notably differ from each other being expressed by
\begin{equation}
\lambda _{\max }^{\left( \nu  \right)} = \frac{{{x_\lambda }}}
{{{x_\nu }}}\;\lambda _{\max }^{\left( \lambda  \right)} \cong
1.76\;\lambda _{\max }^{\left( \lambda  \right)}\quad.
\end{equation}
\noindent For the blackbody temperature of the sun of $T = 5,784\;K$ Eq. (66) provides $\lambda _{\max }^{\left( \lambda \right)} = 500 \; nm$. According to the classification of the visible range by Bruno and Svoronos [2005], the wavelength of the green region is ranging from $495\;nm$ to $570\;nm$. Thus, as stated by Soffer and Lynch [1999], the maximum of the solar radiation is located in the green range when considering the wavelength domain. However, according to Kondratyev [1969] and Petty [2004] the maximum of solar radiation expressed in wavelength units is located in the light blue range ($485\; - \;505\;nm$), rather than in the green range ($505\; - \;550\;nm$). In the case of the other domains the maximum of the solar radiation amounts to a wavelength of $\lambda _{\max }^{\left( {{n_p}} \right)} = \lambda _{\max }^{\left( {{n_s}} \right)} = \lambda _{\max }^{\left( \nu \right)} = 880\;nm$. Since the near-infrared region is ranging from $750\;nm$ to $4\;\mu m$ (e.g., Petty [2004]), the wavelength of the maximum is located in the near-infrared range, as stated by Soffer and Lynch [1999].

\indent For an earth's surface temperature of $T = 300 \; K$ the maximum of the terrestrial radiation amounts to $\lambda _{\max }^{\left( \lambda  \right)} = 9.7\;\mu m$  in the wavelength domain (see Figure 1, upper part). Since the atmospheric window is ranging from $8.3\;\mu m$ to $12.5\;\mu m$ (e.g., M\"{o}ller [1973], Kidder \verb"&" Vonder Haar [1995], Liou [2002], Petty [2004]) that corresponds to "spectroscopic" wave numbers ranging from $1,250\; cm^{-\;1}$ to $800\; cm^{-\;1}$, the value of $\lambda _{\max }^{\left( \lambda  \right)}$ suggests that the atmospheric window envelops this maximum. However, Eq. (67) provides for the maximum of the terrestrial radiation in the other domains: $\lambda _{\max }^{\left( {{n_p}} \right)} = \lambda _{\max }^{\left( {{n_s}} \right)} = \lambda _{\max }^{\left( \nu  \right)} = 17.0\;\mu m$ (see Figure 1, lower part). This means that the maximum of the terrestrial radiation at a temperature of $T = 300 \; K$ is located beyond the atmospheric window. Note that the spectral region of the atmospheric window ranging from  $10\;\mu m$ to $12.5\;\mu m$ is the most common band for meteorological satellites because this part is relatively transparent to radiation upwelling from the earth's surface (Kidder \verb"&" Vonder Haar [1995]).

\indent Nevertheless, either Eq. (62) or Eq. (66) can be applied to determine the ratio ${h/k}$. One obtains ${h/k} = 4.808 \cdot {10^{- \;11}}\;K\;s$. Planck [1901] already considered a formula identical with Eq. (66), but he used $\lambda _{\max }^{\left( \lambda  \right)}\;T \cong 2.94 \cdot {10^{- \;3}}\;m\;K$, in accord with the empirical findings of Lummer and Pringsheim [1900]. Thus, he obtained $h/k = 4.866 \cdot {10^{- \;11}}\;K\;s$. By taking this value and $k^4/h^3 = 1.168 \cdot 10^8\;J\;s^{-\;3}\;K^{- \;4}$ into account, he found: $k = 1.346 \cdot {10^{- \;23}}\;J\;{K^{- \;1}}$ and $h = 6.55 \cdot {10^{- \;34}}\;J\;s$. Obviously, Planck's values only slightly differ from the current ones.

\section{The origin of Planck's constant}

\noindent In his earlier papers on irreversible radiation processes and the entropy and the temperature of radiative heat, respectively, Planck [1900a,b] presented a theoretical derivation of Wien's [1896] radiation law expressed in the wavelength domain $[0,\;\infty]$ by
\begin{equation}
B\left( {\lambda ,\;T} \right) = \frac{{2\;{c^{\,2}}\;b}} {{{\lambda
^5}}}\;\exp \left( { - \;\frac{{a\;c}} {{\lambda \;T}}}
\right)\quad.
\end{equation}
\noindent He only worked with the constant $a = 4.818 \cdot {10^{- \;11}}\;K\;s$ indispensable to make the argument of the in Wien's exponential law non-dimensional, and $b = 6.885 \cdot {10^{- \;34}}\;J\;s$. These values hardly differ from those published by him in 1901. Even though Planck did not define it in these two papers, the constant $b$ is identical with his elementary quantum of action. This means that already in 1896 when Wien published his radiation law a good estimate for the elementary quantum of action could have been obtained (Pais [1995]). Pais [1995], therefore, pointed out that, although Wien did not know this, Wien's exponential law marked the end of the universal validity of classical physics and the onset of a bizarre turn in science.

\indent Using the velocity of light in vacuum,  $c = 3 \cdot {10^8}\;m\;{s^{- \;1}}$, Newton's gravitational constant,  $\gamma = 6.685 \cdot {10^{- \;11}}\;{m^3}\;k{g^{- \;1}}\;{s^{- \;2}}$, and his constants $a$ and $b$ Planck [1900a] introduced natural units of length, mass, time and temperature by (here expressed in SI units)
\begin{equation}\label{7.2}
\sqrt {\frac{{b\;\gamma }} {{{c^3}}}} = 4.13 \cdot {10^{- \;35}}\;m
\quad,
\end{equation}
\begin{equation}
\sqrt {\frac{{b\;c}} {\gamma }} = 5.56 \cdot {10^{- \;8}}\;kg \quad,
\end{equation}
\begin{equation}
\sqrt {\frac{{b\;\gamma }} {{{c^5}}}} = 1.38 \cdot {10^{- \;43}}\;s
\quad,
\end{equation}
\noindent and
\begin{equation}
a\;\sqrt {\frac{{{c^5}}} {{b\;\gamma }}} = 3.50 \cdot {10^{32}}\;K
\quad.
\end{equation}
\noindent He concluded that these quantities keep their natural meaning as long as the laws of gravitation, propagation of light in vacuum, and the first and second principles of thermodynamics stay valid, i.e., even measured by different intelligent beings using different methods, these quantities must again result in the same ones. Recently, the number of fundamental constants in physics was debated by Duff, Okun, and Veneziano [2002]. In contrast to Planck's four fundamental constants, Okun developed the traditional approach with the first three constants, Veneziano argued in favor of at most two (within superstring theory), while Duff advocated zero. The reader is referred to this detailed discussion.

\indent When he presented his improvement of Wien's [1896] exponential law before the Deutsche Physikalische Gesellschaft (German Physical Society) on October 19, 1900 Planck used the form (e.g., Planck [1900c], Kuhn [1978], Landau \verb"&" Lifshitz [1980], Pais [1995], Rechenberg [1995], Giulini [2005])
\begin{equation}
U\left( {\nu ,\;T} \right) = C\;\frac{{{\nu ^3}}} {{\exp \left(
{a\;\frac{\nu } {T}} \right)\; - \;1}}\quad,
\end{equation}
\noindent where C is another constant. As shown by Kramm and Herbert [2006], formula (73) can also be derived by satisfying Ehrenfest's [1911] red requirement that leads to the Rayleigh-Jeans law. In contrast this formula, Eq. (29) gives the form of Planck's radiation law as presented to the German Physical Society on December 14, 1900 and eventually published in 1901 (e.g., Planck [1900d, 1901], Klein [1970], Kuhn [1978], Pais [1995], Rechenberg [1995]). This day may be designated the birthday of quantum theory because the elementary quantum of action and the quantum of energy explicitly occurred in Planck's law (e.g., Pais [1995], Rechenberg [1995]).

\indent Planck's progress in autumn 1900 may be related to his about-face when he gave up some of his opposition to Boltzmann's ideas that results in Planck's use of Boltzmann's connection between entropy and probability. Klein [1970] pointed out that Planck [1900c] had actually begun to consider this connection in his first paper introducing his new distribution, when he referred to the logarithmic dependence of the entropy on the energy implied by this distribution, saying that such a logarithmic dependence was suggested by the theory of probability. It seems plausible. However, Planck's formulation of natural units presented before underlines that he had had a view for physical units. Since his constants $a$ and $b$ have the units $K\; s$ and $J\; s$, respectively, it was surely simple for him to recognize that these two constants can be related to each other by another constant having either the units $K\;{J^{- \;1}}$ or $J\;{K^{- \;1}}$. Like the Boltzmann constant, the latter has the units of the entropy. Planck introduced an additional dimensional constant in his considerations, namely the Boltzmann constant, i.e., $a = {b / k}$ (now $a = {h / k}$). Thus, it is not surprising that the basic relationship Planck referred to was the Boltzmann equation he proceeded to write in the form (Planck [1901])
\begin{equation}
{S_N} = k\;\log \;W\; + \;const.\quad,
\end{equation}
\noindent where the entropy $S_N$ of a set of $N$ oscillators of the same frequency whose total energy is $E_N$ to the quantity $W$ which is proportional to the probability that the set of oscillators have this value for their total energy (Klein [1970]).

\indent We may assume that Planck abandoned his constant $a$ in favor of  $k$, and he determined the latter in his paper of 1901. Consequently, replacing the constant $a$ in the exponential function of Eq. (68) by ${h / k}$ results in the non-dimensional argument ${{h\;\nu } / {\left( {k\;T} \right)}}$, i.e., the ratio of two energies. It seems that Planck's theory presented in his paper of 1901 was formulated in a reverse manner beginning with formula (73).

\indent Planck was not fully satisfied by his derivation. In his Nobel Lecture he stated (Planck [1920]): "However, even if the radiation formula should prove itself to be absolutely accurate, it would still only have, within the significance of a happily chosen interpolation formula, a strictly limited value. For this reason, I busied myself, from then on, that is, from the day of its establishment, with the task of elucidating a true physical character for the formula, and this problem led me automatically to a consideration of the connection between entropy and probability, that is, Boltzmann's trend of ideas; until after some weeks of the most strenuous work of my life, light came into the darkness, and a new undreamed-of perspective opened up before me."

\appendix
\renewcommand{\thesection}{\large Appendix\;\Alph{section}:}
\renewcommand{\theequation}{A\arabic{equation}}
\setcounter{equation}{0}
\section{}\indent The flux density, $F$, and the intensity, $I$, are related to each other by
\begin{equation}
F\; = \;\int\limits_\Omega  {I\;\cos \theta \;d\Omega }\quad,
\end{equation}
\noindent where $\Omega$ is the entire hemispheric solid angle. As the differential solid angle is given by $d\Omega = \sin \theta \;d\theta \;d\varphi$, where  $\theta$ is the zenith angle ($0 \leqslant \theta \leqslant {\pi / 2}$) and $\varphi$ is the azimuthal angle ($0 \leqslant \varphi \leqslant 2\;\pi$), we have
\begin{equation}
F = \int\limits_0^{2\;\pi } {\;\int\limits_0^{{\pi / 2}} {I\left(
{\theta ,\;\varphi } \right)\;\cos \theta \;} } \sin \theta
\;d\theta \;d\varphi
\end{equation}
\noindent or
\begin{equation}
F = \int\limits_0^{2\;\pi } {d\varphi \;\int\limits_0^{{\pi / 2}}
{I\left( {\theta ,\;\varphi } \right)\;\cos \theta \;} } \sin \theta
\;d\theta \quad.
\end{equation}
\noindent It is expressed in units of $J\;{m^{ - 3}}\;{s^{ - 1}}$ or $W\;{m^{ - 3}}$. The flux density is also called the irradiance.

\indent In the case of isotropic radiance, i.e.,  $I$ is independent of $\theta$ and $\varphi$, we have
\begin{equation}
F = I\;\int\limits_0^{2\;\pi } {d\varphi \;\int\limits_0^{{\pi / 2}}
{\cos \theta \;} } \sin \theta \;d\theta \quad.
\end{equation}
\noindent As ${1 / 2}\;\sin \left( {2\;\theta } \right) = \sin \theta \;\cos \;\theta$, Eq. (A4) may also be written as
\begin{equation}
F = \frac{I} {2}\;\int\limits_0^{2\;\pi } {d\varphi
\;\int\limits_0^{{\pi  / 2}} {\sin \left( {2\;\theta }
\right)\;d\theta } }\quad.
\end{equation}
\noindent Defining $x = 2\;\theta $ yields
\begin{equation}
F = \frac{I} {4}\;\int\limits_0^{2\;\pi } {d\varphi
\;\int\limits_0^\pi  {\sin \left( x \right)\;dx} }  = \frac{I}
{2}\;\int\limits_0^{2\;\pi } {d\varphi }  = \pi \;I \quad.
\end{equation}
\noindent Note that blackbody radiance is considered as an example of isotropic radiance.

\vspace{1em}
\noindent
{\bf Acknowledgements :} We are grateful to Dr. Glenn E. Shaw, Professor of Physics emeritus at the University of Alaska Fairbanks, Geophysical Institute, for fruitful discussions and helpful suggestions.

\section*{References}

\begin{description}
\item \textbf{Bohren, C.F.}, \verb"&" \textbf{Clothiaux, E.E.}, (2006): \textit{Fundamentals of Atmospheric Radiation}. Wiley-VCH, Berlin, Germany, 472 pp.
\item \textbf{Boltzmann, L.}, (1877): \"{U}ber die Beziehung zwischen dem zweiten Hauptsatz der mechanischen W\"{a}rmetheorie und der Wahrscheinlichkeitsrechnung respektive den S\"{a}tzen \"{u}ber das W\"{a}rmegleichgewicht.  \textit{Wiener Ber.} \textbf{76}, 373-435 (in German).
\item \textbf{Boltzmann, L.}, (1884): Ableitung des Stefan'schen Gesetzes, betreffend die Abh\"{a}ngigkeit der W\"{a}rmestrahlung von der Temperatur aus der electromagnetischen Lichttheorie. \textit{Wiedemann's Annalen} \textbf{22}, 291-294 (in German).
\item \textbf{Bose, S. N.}, (1924): Plancks Gesetz und Lichtquantenhypothese. \textit{Z. Phys.} \textbf{26}, 178-181 (in German).
\item \textbf{Bruno, T.J.} \verb"&" \textbf{Svoronos, P.D.N.}, (2005): \textit{CRC Handbook of Fundamental Spectroscopic Correlation Charts}. (CRC Press, Boca Raton, FL, 240 pp.
\item \textbf{Chandrasekhar, S.}, (1960): \textit{Radiative Transfer}. Dover Publications, New York, 393 pp.
\item \textbf{D\"{o}ring, W.}, (1973): \textit{Atomphysik und Quantenmechanik: I. Grundlagen.} Walter de Gruyter, Berlin/New York, 389 pp. (in German).
\item \textbf{Duff, M.J.}, \textbf{Okun, L.B.}, \verb"&" \textbf{Veneziano, G.}, (2002): Trialogue on the number of fundamental constants. arXiv:physics/0110060v3.
\item \textbf{Ehrenfest, P.}, (1911): Welche Z\"{u}ge der Lichtquantenhypothese spielen in der Theorie der W\"{a}rmestrahlung eine wesentliche Rolle. \textit{Ann. d. Physik} \textbf{36}, 91-118 (in German).
\item \textbf{Einstein, A.}, (1905): \"{U}ber einen die Erzeugung und Verwandlung des Lichtes betreffenden heuristischen Gesichtspunkt. \textit{Ann. d. Physik} \textbf{17}, 164-181 (in German).
\item \textbf{Einstein, A.}, (1917): Zur Quantentheorie der Strahlung. \textit{Physik. Zeitschr.} \textbf{XVIII}, 121-128 (in German).
\item \textbf{Einstein, A.}, (1924): Quantentheorie des einatomigen idealen Gases. Sitzungsber. Preuß. Akad. Wiss., Berlin, 262-267; \textit{Zweite Abhandlung}. Sitzungsber. Preuß. Akad. Wiss., Berlin, 3-14 (1925); \textit{Quantentheorie des idealen Gases}. Sitzungsber. Preuß. Akad. Wiss., Berlin, 18-25 (1925, in German; as cited by Rechenberg, 1995).
\item \textbf{Feynman, R.P., Leighton, R.B.}, \verb"&" \textbf{Sands, M.}, (1963): \textit{The Feynman Lectures on Physics - Vol. 1}. Addison-Wesley Publishers, Reading, MA/Palo Alto/London.
\item \textbf{Gerlich, G.}, \verb"&" \textbf{Tscheuschner, R.D.}, (2009): Falsifcation of the atmospheric CO2 greenhouse effects within the frame of physics. \textit{Int. J. Mod. Phys.} \textbf{B23} 275-364 (see also http://arxiv.org/abs/0707.1161v4).
\item \textbf{Giulini, D.}, (2005): \textit{Es lebe die Unverfrorenheit! Albert Einstein und die Begr\"{u}ndung der Quantentheorie}. arXiv:physics/0512034.
\item \textbf{Gooody, R.M.} \verb"&" \textbf{Yung, Y.L.}, (1989): \textit{Atmospheric Radiation}. Oxford University Press, New York/Oxford, 519 pp.
\item \textbf{Jeans, J. H.}, (1905): On the laws of radiation. \textit{Proc. Royal Soc. A} \textbf{76}, 546-567.
\item \textbf{Kidder, S.Q.}, \verb"&" \textbf{Vonder Haar, T.H.}, (1995): \textit{Satellite Meteorology}. Academic Press, San Diego/New York/Boston/London/Sydney/Tokyo/Toronto, 466 pp.
\item \textbf{Kirchhoff, G.}, (1860): Ueber das Verh\"{a}ltniss zwischen dem Emissionsverm\"{o}gen und dem Absorptionsverm\"{o}gen für W\"{a}rme und Licht. \textit{Ann. d. Physik u. Chemie} \textbf{109}, 275-301 (in German).
\item \textbf{Klein, M.J.}, (1970): \textit{Paul Ehrenfest - Volume 1: The Making of a Theoretical Physicist}. North-Holland Publishing Comp., Amsterdam/London, 330 pp.
\item \textbf{Kondratyev, K.YA.}, (1969): \textit{Radiation in the Atmosphere}. Academic Press, New York/London, 912 pp.
\item \textbf{Kramm, G.}, \verb"&" \textbf{Herbert, F.}, (2006): Heuristic derivation of blackbody radiation laws using principles of dimensional analysis. \textit{J. Calcutta Math. Soc.} \textbf{2} (2), 1-20.
\item \textbf{Kuhn, T.S.}, (1978): \textit{Black-Body Theory and the Quantum Discontinuity 1894-1912}. Oxford University Press, Oxford/New York, 356 pp.
\item \textbf{Landau, L.D.} \verb"&" \textbf{Lifshitz, E.M.}, (1980):
\textit{Course of Theoretical Physics - Vol. 5: Statististical Physics}. Pergamon Press, Oxford/New York/Toronto/Sydney/Paris/Frankfurt, 544 pp.
\item \textbf{Lenoble, J.}, (1993): \textit{Atmospheric Radiative Transfer}. A. Deepak Publishing, Hampton, VA,, 532 pp.
\item  \textbf{Liou, K.N.}, (2002): \textit{An Introduction to Atmospheric Radiation}. Academic Press, Amsterdam/Boston/London/New York/Oxford/Paris/San Diego/San Francisco/Singapore/ Sydney/Tokyo, 577 pp.
\item \textbf{Lorentz, H. A.}, (1903): On the emission and absorption by metals of rays of heat of great wave-length. \textit{Proc. Acad. Amsterdam} \textbf{5}, 666-685.
\item  \textbf{Lummer, O.} \verb"&" \textbf{Pringsheim, E.}, (1900): \"{U}ber die Strahlung des schwarzen K\"{o}rpers f\"{u}r lange Wellen. Verh. d. Deutsch. Phys. Ges. \textbf{2}, 163-180 (in German).
\item \textbf{Modest, M.F.}, (2003): \textit{Radiative Heat Transfer}. Academic Press, Amsterdam/Boston/ London/New York/Oxford/Paris/San Diego/San Francisco/Singapore/Sydney/Tokyo, 822 pp.
\item \textbf{M\"{o}ller, F.}, (1973): \textit{Einf\"uhrung in die Meteorologie}, \textit{Bd. 2: Physik der Atmosph\"{a}re}. Bibliographisches Institut, Mannheim/Wien/Zürich, 223 pp. (in German).
\item \textbf{Paschen, F.}, (1896): Ueber die Gesetzm\"{a}{\ss}igkeiten in den Spectren fester K\"{o}rper. \textit{Ann. d. Physik} \textbf{58}, 455-492 (in German).
\item \textbf{Pauli, W.}, (1973): \textit{Statistical Mechanics} (Vol. 4 of Pauli Lectures on Physics). MIT Press, Cambridge, MA, 173 pp.
\item \textbf{Peixoto, J.P.} \verb"&" \textbf{Oort, A.H.}, (1992): \textit{Physics of Climate}. Springer, New York/Berlin/ Heidelberg, 520 pp.
\item \textbf{Petty, G.W.}, (2004): \textit{A First Course in Atmospheric Radiation}. Sundog Publishing, Madison, WI, 444 pp.
\item \textbf{Pais, A.}, (1995): Introducing Atoms and their Nuclei. In: \textbf{Brown, L.M.}, \textbf{Pais, A.}, and \textbf{Sir Pippard,
B.} (eds.), \textit{Twentieth Century Physics - Vol. I}. Institute of Physics Publishing, Bristol and Philadelphia, and American Institute of Physics Press, New York, pp. 43-141.
\item \textbf{Planck, M.}, (1900a): \"{U}ber irreversible Strahlungsvorg\"{a}nge. \textit{Ann. d. Physik} \textbf{1}, 69-122 (in German).
\item \textbf{Planck, M.}, (1900b): Entropie und Temperatur strahlender W\"{a}rme. \textit{Ann. d. Physik} \textbf{1}, 719-737 (in German).
\item \textbf{Planck, M.}, (1900c): \"{U}ber eine Verbesserung der Wien'schen Spektralgleichung. \textit{Verh. d. Deutsch. Phys. Ges.} \textbf{2}, 202-204 (in German).
\item \textbf{Planck, M.}, (1900d): Zur Theorie des Gesetzes der Energieverteilung im Normalspectrum. \textit{Verh. d. Deutsch. Phys. Ges.} \textbf{2}, 237-245 (in German).
\item \textbf{Planck, M.}, (1901): Ueber das Gesetz der Energieverteilung im Normalspectrum. \textit{Ann. d. Physik} \textbf{4}, 553-563 (in German).
\item \textbf{Planck, M.}, (1920): The genesis and present state of development of the quantum theory. Nobel Lecture, June 2, 1920.
\item \textbf{Lord Rayleigh} (J. W. Strutt), (1900): Remarks upon the law of complete radiation. \textit{Phil. Mag.} \textbf{49}, 539-540.
\item \textbf{Lord Rayleigh}, (1905): The dynamical theory of gases and of radiation. \textit{Nature} \textbf{72}, 54-55.
\item \textbf{Rechenberg, H.}, (1995): Quanta and Quantum Mechanics. In: \textbf{Brown, L.M.}, \textbf{Pais, A.}, and \textbf{Sir Pippard, B.} (eds.), \textit{Twentieth Century Physics - Vol. I}. Institute of Physics Publishing, Bristol and Philadelphia, and American Institute of Physics Press, New York, pp. 143-248.
\item \textbf{Rubens, H.} \verb"&" \textbf{Kurlbaum, R.}, (1900): \"{U}ber die Emission langwelliger W\"{a}rmestrahlen durch den schwarzen K\"{o}rper bei verschiedenen Temperaturen. Sitzungsber. d. K. Akad. d. Wissensch. Berlin vom 25. Oktober 1900, 929-941 (in German).
\item \textbf{Rubens, H.} \verb"&" \textbf{Kurlbaum, R.}, (1901): Anwendung der Methode der Reststrahlen zur Pr\"{u}fung des Strahlungsgesetzes. \textit{Ann. d. Physik} \textbf{4}, 649-666 (in German).
\item \textbf{Rybicki, G.B.} \verb"&" \textbf{Lightman, A. P.}, (2004): \textit{Radiative Processes in Astrophysics}. Wiley-VCH, Weinheim, Germany, 400 pp.
\item \textbf{Semat, H.} \verb"&" \textbf{Albright, J.R.}, (1972): \textit{Introduction to Atomic and Nuclear Physics}. Holt, Rinehart \verb"&" Winston, Inc., New York/Chicago/San Francisco/Atlanta/Dallas/Montreal/ Toronto/London/ Sydney, 712 pp.
\item \textbf{Soffer, B.H.}, \verb"&" \textbf{Lynch, D.K.}, (1999): Some paradoxes, errors, and resolutions concerning the spectral optimization of human vision. \textit{Am. J. Phys.} \textbf{67}, 946-953.
\item \textbf{Stefan, J.}, (1879): \"{U}ber die Beziehung zwischen der W\"{a}rmestrahlung und der Temperatur. \textit{Wiener Ber. II}, \textbf{79}, 391-428 (in German).
\item \textbf{Wien, W.}, (1894): Temperatur und Entropie der Strahlung. \textit{Ann. d. Physik} \textbf{52}, 132-165 (in German).
\item \textbf{Wien, W.}, (1896): Ueber die Energieverteilung im Emissionsspectrum eines schwarzen K\"{o}rpers. \textit{Ann. d. Physik} \textbf{58}, 662-669 (in German).
\end{description}

\vspace{1em}
\footnoterule
\vspace{.5em}
\noindent
$^1$UNIVERSITY OF ALASKA FAIRBANKS, GEOPHYSICAL INSTITUTE \\
903 KOYUKUK DRIVE, P.O. BOX 757320\\ FAIRBANKS, ALASKA 99775-7320\\
UNITED STATES OF AMERICA \\
TEL.: +1 907 474 5992, FAX : +1 907 474 7290\\
E-MAIL : kramm@gi.alaska.edu

\noindent
$^1$UNIVERSITY OF ALASKA FAIRBANKS, GEOPHYSICAL INSTITUTE \\
$^2$UNIVERSITY OF ALASKA FAIRBANKS, COLLEGE OF NATURAL SCIENCE AND MATHEMATICS, \\
DEPARTMENT OF ATMOSPHERIC SCIENCES \\
903 KOYUKUK DRIVE, P.O. BOX 757320\\ FAIRBANKS, ALASKA 99775-7320\\
UNITED STATES OF AMERICA \\
TEL. : +1 907 474 7910, FAX : +1 907 474 7290\\
E-MAIL : molders@gi.alaska.edu
\end{document}